# Sparse sampling: Spatial design for monitoring stream networks [*],[†]

### Melissa J. Dobbie, Brent L. Henderson

*CSIRO Mathematical and Information Sciences, Australia*
*e-mail:* melissa.dobbie@csiro.au
*e-mail:* brent.henderson@csiro.au

and

### Don L. Stevens, Jr

*Oregon State University, USA*
*e-mail:* stevens@stat.oregonstate.edu

**Abstract:** Spatial designs for monitoring stream networks, especially ephemeral systems, are typically non-standard, 'sparse' and can be very complex, reflecting the complexity of the ecosystem being monitored, the scale of the population, and the competing multiple monitoring objectives. The main purpose of this paper is to present a review of approaches to spatial design to enable informed decisions to be made about developing practical and optimal spatial designs for future monitoring of streams.

## 1. Introduction

Aquatic ecosystems are of critical international importance, contributing substantial environmental, ecological, social and economic value. Many of these systems are under pressure from anthropogenic sources such as increasing development, land use practices, contamination of surface and ground water and the atmospheric deposition of pollutants. Human-induced climate change is also increasingly affecting those ecosystems.

In order to identify changes and demonstrate the effectiveness of any management interventions undertaken in reducing those pressures, there is a strong focus on ascertaining and reporting on ecosystem health to inform ecosystem managers and the wider community. These assessments often need to be made across large spatial regions such as catchments or jurisdictional boundaries and depend on monitoring that is underpinned by a spatial design.

Our focus is on large-scale monitoring designs for freshwater streams, though many of the conclusions will follow for lakes, estuaries and other water domains. Often multiple indicators of ecosystem health are of interest; these might include traditional physical and chemical water quality variates, biological indicators for macroinvertebrates, fish or macrophytes, as well as terrestrial indicators such as riparian vegetation classification and extent. Spatial designs for monitoring are appropriate for any measure of ecosystem health that may be referenced spatially. As such the spatial design is developed in the context of the population of interest such as a stream network and is driven by the monitoring and reporting objectives of the study, not the indicators that we are monitoring in that population.

---

[*]This is an original survey paper.
[†]This paper was accepted by Noel Cressie, Associate Editor for the IMS.

Numerous statistical challenges are faced in developing large-scale monitoring programs for assessing ecosystem health. The development of practical and appropriate "sparse" spatial designs is a key challenge. The description of sampling for these applications as sparse is motivated by the tension between the cost of sampling and number of sites to be sampled; the allocation of resources across space and time; and, in the context of freshwater applications, the unique structure of the target population within the landscape.

We acknowledge that the idea of sparse sampling is not new - most sampling methods seek response from a fraction of a population at best - but in the context of designing monitoring programs of stream networks it helps capture the essence of important design issues that arise and need to be considered in more detail.

As with most large-scale sample design problems, the central challenge is how to allocate sampling resources across space (and time) to maximize the information available, which can then be used to make reliable and credible inferences or predictions about the response(s) of interest. Of course, as with any large-scale monitoring program, it is essential that the sampling or statistical design and the response (or operational) design are considered simultaneously to meet stringent resource constraints.

Our emphasis here is on the sampling or statistical design, and in particular on the spatial aspects of that design, which we call the *spatial design.* We will present a synthesis of spatial design research, particularly in relation to monitoring of stream networks, and document the current status of this area of research with the ultimate aim of informing development of appropriate spatial designs for future monitoring of stream networks.

Spatial designs for monitoring stream networks are typically non-standard and can be very complex, reflecting the complexity of the system being monitored. Whilst progress has been made in research and application of spatial designs for monitoring natural resources, especially in the United States of America (e.g. [88]; [90]), there is limited published knowledge and application of spatial designs for monitoring stream condition elsewhere. Section 2 provides a review of spatial sampling design including descriptions of the two most common approaches, model-based and probability-based sampling. The strengths and limitations of some of the more standard designs for applications in monitoring natural resources such as stream networks are discussed, and a review of spatial balance, an important attribute of a spatial sampling design, and some of the designs that feature spatial balance are also provided.

Historically, professional judgement and opinion has played an important role in monitoring design of aquatic systems such as stream networks. In Section 3 we take a look at ways of capturing this source of information (i.e. through expert elicitation) and its potential role and use in informing the design.

Quantifying trends over time is often a key management objective of large-scale environmental monitoring programs, and we consider methods for incorporating a temporal sampling strategy into the overall spatial design in Section 4. In Sections 5 and 6 we briefly discuss other important design considerations and some technological advances of relevance to this research area, respectively.

A summary of the design considerations for some key stream condition monitoring programs are presented in Section 7, followed by concluding remarks in Section 8.

## 2. Review of spatial sampling design

### 2.1. Sampling environmental resource populations

The design of sample surveys is a mature area in the statistical literature. Cochran [14] and Sarndal et al. [73] provide excellent accounts of the essential attributes of good sampling design, including clear definitions of the population of interest, the sample units, the sample frame and how the sample is to be drawn.

Designs for environmental monitoring however bring with them additional challenges. Environmental resources are often expensive, time-consuming and highly complex systems to sample. They may also change or move over time and/or space, with streams being less dynamic than air but more dynamic than soil. Furthermore, a reliable sampling frame for that target population is not always easy to produce, and it is typically only possible to sample a small proportion of that population.

The size of the population will impact on the spatial design both in theory and in practice. The ramifications of large-scale (such as landscape-scale) environmental populations on a spatial design are numerous and include:

- sparser representation of that population through sample sites selected via the spatial design;
- the potential inability for the sample to adequately represent the population;
- the likelihood that the population is spatially heterogenous which should be controlled through appropriate approaches to harness the large-scale variation present;
- increased non-stationarity of the key underlying processes; and
- practical challenges in meeting tight time and resource constraints.

There have been several published reviews of spatial design approaches to sampling environmental populations. Stein & Ettema [83] present a review of spatial sampling procedures for developing optimal spatial designs for comparison of ecosystems. Whilst their review is undertaken within the broad context of ecological, environmental and agricultural environments, they present an application of spatial constrained sampling of soil-biodiversity in grasslands.

Dixon & Chiswell [22] provide a broad review of aquatic monitoring program design, covering aspects such as information goals, data and trend analysis, types of designs, indicators, the spatial distribution of sampling sites, the sampling frequency and the cost-effectiveness of designs. Maher et al. [50] also discuss a framework for designing sampling programs of natural resources including some discussion on the spatial selection of sampling sites. The book edited by Sanders, "Design of Networks for Monitoring Water Quality" ([71]), is one of few books written on the specific topic and application.

Stevens & Olsen [88] describe some of the attributes of environmental resources that affect environmental monitoring design. They describe how environmental monitoring designs involve sampling populations of different dimensions namely 0-dimensional (points; e.g. trees, small lakes), 1-dimensional (lines; e.g. rivers and streams) and 2-dimensional (areas; e.g. lakes, estuaries, wetlands) sampling units. They also emphasise that environmental monitoring designs are often needed to sample a continuous population, e.g. a stream, where individual sampling units are not as obvious or clearly defined as in other types of sample survey. This issue extends to the definition of reliable sampling frames for such monitoring. The result is that the units selected may differ substantially from units of interest. In the broader context of monitoring aquatic systems, typically sample site selections produce a "location" and specific units of interest are determined on-site.

The response of interest may exhibit spatial patterns and structure. For example there may be spatial gradients or periodicity. Winkels & Stein [110] designed an optimal sampling design for monitoring contaminated sediments in a lake, which required them to understand the spatial variability of the parameters of interest over the population. Stratification may be a good way to account for spatial structure in the response, but appropriate strata can be difficult to choose and may not persist over time. For example, strata based on regional or jurisdictional boundaries may necessarily change with political decisions.

There may be many environmental responses of interest that are inter-dependent (e.g. different water quality indicators). Appealing to knowledge of these correlations may help inform the design. These may also be subject to substantially more measurement error than routinely found in other types of surveys.

Non-response in environmental monitoring can be substantial. It may not be possible to obtain responses from a substantial proportion of sampling locations for reasons such as ease of physical access, safety or permission. In stream network surveys for assessing ambient condition, streams that flow only in direct response to precipitation, otherwise known as ephemeral streams, may also be a contributor to non-response.

Methods for sampling environmental resources have often been fairly ad hoc. Convenience sampling has often been used, and appeals to expert knowledge to choose sample locations with easy access or that may be originally useful for other purposes. A key complication of convenience sampling that arises is that the relationship between the sample data and population characteristics of interests is not known, and the basis for extrapolation and inference is therefore necessarily unclear.

Representative sampling is also popular in sampling environmental resource populations and relies on the selection of sites that are indicative of the region/resource of interest. This appeals to expert knowledge but is often challenged on the grounds that the site selection is inherently subjective, that sites which are representative for one variable may not be representative for any other variable, and because if sites are truly representative of average response then the extremes will be suppressed.

Statistical considerations will always represent only one aspect of a sampling design. There are many practical and scientific considerations which impact on the design, particularly for sampling of stream networks. Indeed, convenience sampling is often used in the field for these reasons. However, a statistical perspective is important and fundamental at the design stage for ensuring credibility, reliability and validity of information collected. In support of this, Olsen et al. [64] state that statisticians

- can ensure that the information is gathered in a scientifically defensible manner and that stated objectives will be met in a way that is cost-effective;
- will take the whole monitoring framework into account and thus bring a big picture perspective to the problem;
- may assist in "borrowing strength" (i.e. through use of historical data) from existing designs or suggest modifying existing designs to make them more efficient and improve information to decision-makers;
- can improve decision-making by providing estimates of uncertainty about condition or trend at scales of relevance to the design and objectives of the program; and
- can provide extension from knowledge garnered about sites to an entire region, which is often non-trivial.

The focus of this report is primarily on the statistical aspects associated with selecting an appropriate spatial sample.

Stehman [81] suggested the following criteria will produce a good survey design in an environmental application:

- low estimated variance
- uncertainty in estimated variance is calculable
- spatial balance
- simplicity, and
- cost-effectiveness.

Theobald et al. [90] added flexibility to this list, which is a particularly important criterion for aquatic monitoring as non-response can potentially be substantial. The general principles of a good experimental design - the 5 R's, namely randomization, replication, reference (or control as it's commonly known in traditional experimental design), relevance and repeatability - should also be adhered to as much as possible in order to ensure sound and credible inferences may be made from the data that is collected.

## *2.2. Spatial monitoring design*

There are three popular statistically-based philosophies for choosing the spatial monitoring design:

1. **Geometric** approaches are typically based on heuristic arguments and include regular lattices, triangular networks, or space-filling designs. These approaches are typically used for exploratory purposes e.g. Muller [57].
2. **Probability-based** approaches select sites via a probability sample and use survey sample methods to make inferences about the population of interest or some characteristic of it.
3. **Model-based** approaches base the inference about the target population on an explicit specification of the relationship between the selected sites and the population in terms of a statistical model. Model-based design can also include implicit model selection methods such as targeted sampling, representative sites and convenience sampling, which were previously described in Section 2.1 and will not be discussed further in this review.

There is considerable discussion of the contrasts between model-based and design-based inference in the literature (e.g. [29]; [72]). Brus & De Gruitjer [6, 7] and De Gruitjer & Ter Braak [19] focus on the differences in spatial inference. Theobald et al. [90] provide a concise description of the differences in relation to environmental monitoring. Table 1 provides a summary of each mode of inference and details many of these contrasts.

As a compromise between these two broad strategies, Cressie et al [17] discuss design for data collection in ecological studies from a statistical modelling perspective. They show how probability-based sampling designs can be incorporated into statistical models, resulting in what is termed model-assisted design-based inferences. Further discussion about this form of sampling is presented in Section 2.4.7.

The most appropriate spatial monitoring design approach in a specific circumstance is intimately linked to the objectives of the monitoring program. The design necessary for making reliable spatial predictions in a region may be quite different to the design required to report on a distributional quantity such as the mean for a region. It is essential that there is clarity of purpose if we want to make informed decisions about the spatial design. We now consider each of these approaches, particularly probability-based and model-based designs, in more detail.

## *2.3. Geometric design*

Geometric approaches consider how well a set of design points covers the domain. There is no dependence on the spatial covariance or the stochastic model. The design criterion is based on geometry and the distance between both current and potential sample locations. Royle & Nychka [70] and Nychka & Salzman [61] describe space-filling designs. Dixon et al. [23] describe an approach for the selection of river sampling sites that considers potential locations in a river network and uses simulated annealing to choose an optimum configuration according to a cost function that is based on cumulative information at site locations (e.g. upstream area, flow) and the expected cost of subsequent investigations to determine the source of a pollutant. Geometric methods are not described any further in this paper.

Table 1
*Model-based versus design-based inference.*

| **Model-based inference** | **Design-based inference** |
|---|---|
| • Uses model to describe relationship between sample and population | • Probability-based selection method |
| • Enables very general and precise inference from limited sample | • Estimation, testing and prediction are based on the inclusion probabilities |
| • Valuable when interest in predictions at individual points | • Generality and validity comes from the probability-based sampling |
| • Inference borrows strength from model | • Inference from a properly executed design is compelling |
| • High spatial correlations may be important. If correlations are weak there may be limited ability to 'borrow strength' from nearby observations in making inferences | • Design-based methods gives estimates of characteristics or attributes of the real target population e.g. its mean, variance or distribution |
| • Precision of the inference is judged relative to the model. Issues include appropriateness of model and inability to check that the model is reasonable | • Design-based methods result in estimates of population characteristics that are unbiased and have objective assessments of uncertainty |
| • Randomisation is embedded in the underlying process rather than in the selection of monitoring sites | • Assumes it is feasible to sample randomly |
| • Some ecological systems may demand complex models | • Probability sampling allows estimates to be aggregated from the local to the national level (e.g. for hierarchical scales of reporting) |
| | • Can incorporate prior knowledge and understanding in both the design and analysis phases |

### 2.4. *Probability-based design*

Probability-based designs assume a fixed underlying process and use probability sampling to select the monitoring sites. This contrasts with model-based designs, where the stochastic element is embedded in the model process. The use of probability sampling is critical to design-based inference. A probability sampling design for an explicitly-defined resource population is a means to certify that the data collected are free from any selection bias, conscious or not. A probability sampling design has three distinguishing features:

- the population being sampled is explicitly described;
- every element in the population has the opportunity to be sampled with known probability; and

- the selection is carried out by a process that includes an explicit random element.

These features provide mathematical foundations for statistical inference. Randomisation is particularly important as it avoids bias and ensures the sample is representative.

Probability-based sampling designs include:

- simple random sampling (SRS)
- systematic sampling (SyS)
- stratified sampling, random or systematic
- two-stage or cluster sampling
- double sampling
- adaptive sampling, and
- spatially-balanced sampling.

Overton [65] emphasizes probability sampling in the context of ecological monitoring, whereas Gilbert [28] provides a theoretical account and illustration of most of these designs in an environmental context. Figure 1 gives a one-dimensional illustration of some of these designs, motivated by the linear nature of rivers. It was reproduced (with permission) and adapted from a figure in Gilbert that compares designs in a similar manner.

The primary focus of probability-based sampling is to enable us to make inferences about a relevant attribute for a population on some spatial domain $D$. As an example, for a stream network we might be interested in determining the length of the network that is of a particular condition (e.g. impaired). Design-based samples typically seek to enable inference about a feature of the population attribute such as a mean, total, variance, proportion or a distribution function.

Some of the common probability-based design approaches are now defined and the advantages and disadvantages of applying them are discussed.

### 2.4.1. Simple random sampling

As the name infers, simple random sampling (SRS) is the simplest form of probability sampling where a series of random locations $(x, y)$ is generated from a population under no conditions. Advantages of this design include simple prescriptive calculations for the mean response at those locations and thus straightforward statistical inferences, and flexibility to increase or decrease the sample size if required. On the downside, as the locations are treated as independent, the variance of the response can be high due to some parts of the region being represented more heavily than others due to chance (i.e. sites tend to be clustered). In addition, the design does not allow for adjustments for spatial autocorrelation, which may arise through spatial structure captured by important covariates. SRS provides no assurance of spatial balance or regularity so SRS samples are frequently inefficient at covering the space. For monitoring stream condition, it is an impractical and inefficient way to sample.

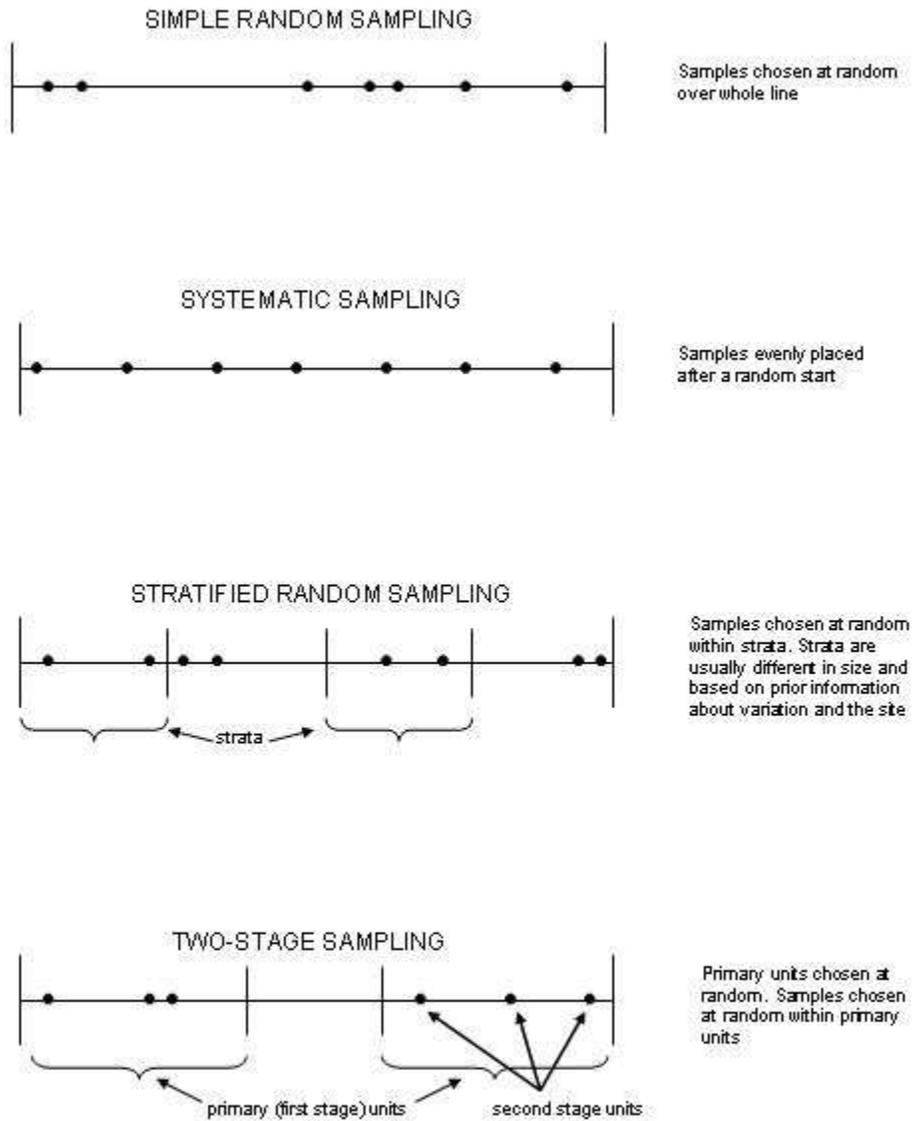

Fɪɢ 1. *Illustration of simple random sampling, systematic sampling, stratified random sampling and two-stage sampling in one dimension so there are clear analogies to sampling in linear structures such as rivers. This figure has been reproduced (with permission) and adapted from a figure that appears in Gilbert [28].*

## 2.4.2. Systematic sampling

Systematic sampling involves taking samples at locations according to some regular pattern. For instance, systematically selecting locations from a regular grid overlaid on the target population will ensure they are equally spaced. This sampling approach also ensures good spatial balance. Furthermore, it is more precise than a simple random sample if there is spatial autocorrelation because the tendency for sampling locations to clump will reduce the efficiency of the design. The broad coverage of systematic samples can be an advantage when sampling rare and clustered populations.

Randomness may be introduced to the design by selecting a random start (i.e. a random location for a 1-dimensional population such as a river). This gives every location a chance of selection though some locations will always be included together. Multiple starts (or placements) are advantageous as they reduce the potential impact of patterned or periodic response, and they permit a purely design-based variance estimate.

Systematic sampling methods may be challenged by populations exhibiting patterned or periodic responses. One classic example of this for an aquatic context arises when weekly compliance samples for a waste water treatment plant are taken consistently on a Monday morning despite knowledge of a likely weekend effect, e.g. reduced flushing or outflow compared to that recorded for a midweek sample. Also, systematic sampling may be difficult to apply for some finite populations, e.g. lakes in a particular region. There is typically less flexibility to change the sample point density or add new points without updating the whole design. Systematic sampling may not adequately sample less common sub-populations unless sample sizes are very large. It can be difficult to optimise the design for different costs of access or to accommodate variable inclusion probability, though stratified systematic sampling may offer a partial solution.

To avoid issues with response variables of interest being strongly dependent (e.g. such as nutrients), Caeiro et al. [10] adopt a systematic unaligned sampling design ([4]) for spatial sampling in an estuarine system. They found that this approach avoids the periodicities of the systematic approach, gives good coverage over an area, is efficient, and deals with most distributions.

Inference for the population under systematic sampling is straightforward, though standard errors are not as well supported because there is a need to make up for the lack of randomness in the location selection by assuming the spatial process is random.

The random-tessellation stratified (RTS) design (e.g. [66]) is a compromise between systematic and simple random sampling that resolves issues concerning periodic or patterned responses not handled by systematic sampling designs. Specifically, RTS samples are selected by randomly locating a regular grid (e.g. square or triangular) over the population domain with spacing chosen to provide the required spatial resolution. Random points from within each random tessellation cell are selected to form the required sample. Like systematic designs, RTS designs do not allow variable probability spatial sampling. There are also issues of applying RTS designs to finite (e.g. lakes) and linear (e.g. rivers) resources and when spatial clustering occurs in the target population. Olsen et al. [63] discuss application of global grids to large-scale environmental sampling of natural resources, one of which is the RTS design.

### 2.4.3. Two-stage sampling

Two-stage sampling (or multi-stage sampling) is useful when the target population can be divided into a collection of 'primary units'. The first stage of the sampling involves taking a random sample of primary units. For each of these primary units, a second stage of sampling is applied and sample units selected randomly or systematically. Two-stage sampling is often necessary for practical reasons. As an example, a set of monitoring sites may be chosen randomly in some way from a list of possible monitoring sites. For each of these sites several subsamples of water may be taken for physical, chemical and biological analysis. In this case the selection of the sites represents the first stage and the water subsamples represent the second stage of the sampling.

Cluster sampling is closely related to two-stage sampling. The first stage, and selection of the primary unit, is identical. The difference lies in the secondary stage, where units in the cluster are not selected randomly or systematically, but rather all units in the cluster are typically sampled. Cluster sampling is usually necessary when it is difficult to sample the individual units in the population.

### 2.4.4. Extending these basic designs

Monitoring objectives may often include requirements that basic designs can not address efficiently. For instance, estimates for particular subpopulations may require greater sampling effort due to administrative restrictions or operational costs, or the presence of the environmental or ecological resource may be restricted to particular habitats. Some elements of the population may be more important to sample than others, exhibit higher uncertainty, or correspond to locations that are more difficult (i.e. due to time or money) to access than others. For example, in a lake survey we may wish to sample lakes in regions more susceptible to drought more intensively. There are two popular approaches to dealing with some of these complexities: (i) *stratification*, where for instance we may wish to sample larger lakes with a higher probability because they are less numerous or contain more of the total volume, and (ii) *variable probability sampling*, where we might choose to sample lakes according to probabilities that may be proportional to the size of the lake. We now describe each of these approaches in turn.

### 2.4.5. Stratified sampling

A stratified sampling design divides the population into a collection of strata that exhaustively cover the region of interest. These strata may represent different sub-regions or sub-populations of interest and thus help describe important spatial structure in the population. They may also reflect operational or administrative aspects (e.g. local government boundaries) of the region. Measurements within strata should typically be more homogenous than measurements across the whole population. The largest gains in efficiency over simple random or systematic sampling accrue when the variability between strata is large compared to the variation within the strata.

Stratification allows us to allocate sampling resources to the different strata at different rates. This may be useful for estimating less common sub-populations and incorporating costs into the design (e.g. we might put more effort into areas that are accessible and less costly to sample). It also provides us with a way of providing some spatial balance/representation by requiring samples be taken within each stratum. Stratification makes use of prior knowledge of the population and the spatial characteristics of the region. The strata do however need to be consistent through time in order to reliably demonstrate any temporal changes and trends. This emphasises the importance of choosing strata at the outset for long term studies so that they will persist over time.

The choice of the number of strata and the number of samples to be taken within each stratum is a complex question and depends intimately on the variability associated with each strata and the relative importance of that strata or sub-population. Non-response in environmental sampling is a challenging problem but can be overcome by "over-sampling", that is by selecting more samples than originally intended. If too few samples are originally selected and there is some non-response, then it may not be possible to estimate the variance associated with that strata.

Stratified sampling is more complex than simple random or systematic sampling but the inference is well-formed and widely accessible (see for instance [14] or [28]). In the context of monitoring streams in a network, strata are often formed using specific classifications of the network such as stream order or type.

Stream order is a measure of the position of a stream within a river network and it is often thought of as reflecting the relative size of the stream within the network. There are various ways of defining stream order, the most common being attributed to Strahler [89] and Shreve [76]. Strahler's stream order system is a simple method of classifying stream segments based on the number of tributaries upstream. A stream with no tributaries (headwater stream) is considered a first order stream, whilst a segment downstream of the confluence of two first order streams is a second order stream. Thus, a $n$th order stream is always located downstream of the confluence of two $(n-1)$th order streams. Shreve's stream order system is similar although when two streams converge, the stream downstream of the confluence is assigned an order equal to the sum of the immediate upstream orders. Stream type may refer to the geographical characteristic of the stream such as upland, lowland, wallum, etc.

Liebetrau [46] provides a good example of stratified sampling of a stream network. The sampling unit is stream links, which is typically a section of a stream between two confluences. Stratified sampling is used to sample from different stream orders with inclusion probabilities that weight the importance of sampling that order. For example, the largest (Shreve) stream order (58) might be sampled with probability 1 because it is essential it be included, being the only stream segment of that order, while first order streams might have a small inclusion probability given they are numerous in the population. Other stratification variables that might have been used instead of (or as well as) stream order include drainage area, mean annual flow and total upstream path length.

### *2.4.6. Variable probability sampling*

Variable probability sampling is a generalisation of stratified sampling that allows selection probabilities to vary continuously, instead of being constant within discrete strata and only varying among strata. In a variable probability design of fixed sample size, the inclusion of an element $u$ in a sample will occur with inclusion probability $\pi_u$ that is proportional to an auxiliary variable $z_u$ that is measured on element $u$. An example of an auxiliary variable in the context of monitoring a stream network is the contributing catchment area upstream of each stream segment. In a forestry context, if we were interested in estimating the total volume of a forest, a natural auxiliary variable would be the basal circumference of the tree as that might ensure we concentrate our sampling efforts on the largest trees. Variable probability sampling designs may be generalised to include more than one auxiliary variable. For instance in the forestry context we might use the basal circumference and the height of the tree. Most importantly, auxiliary variables are useful for discriminating between sub-populations of interest so that variable probability sampling can be used to allocate samples to the sub-populations accordingly, and thus improve the precision of the results. Diaz-Ramos et al. [20] note that population estimators will tend to have a lower variance if the attribute of interest and the auxiliary variable $z_u$ are strongly positively correlated.

Variable probability offers greater flexibility, although there are obvious complexities to be considered in practice. The cautions on not changing strata also apply to selection probabilities, which should not be changed, even if conditions change, in order to develop reliable inference about any temporal change.

The Horvitz-Thompson Theorem ([32]) or its continuous analogue ([84]) typically underly the estimation methods used for data generated by probability-based designs. For a continuous linear case such as a river network, suppose that $y(x)$ defines a fixed but unknown attribute at location $x$ in spatial domain $D$. Suppose for illustration that our interest is in an estimate of the total of that attribute, i.e.

$$y_T = \int_D y(x) \ dx.$$

Other quantities that might be considered are means, variances, proportions or distributions. Consider a sample of locations $x_1, \ldots, x_n$ chosen from a universe $U$ that contains the spatial domain $D$ as a subset, according to a known inclusion probability distribution. Suppose that $y(x_1), \ldots, y(x_n)$ denotes attribute values, determined for each of the sample locations within $D$. Assume that $f(x_1, \ldots, x_n)$ is the joint probability density function (pdf) of the sample locations, $f_i(x)$ is the marginal pdf of location $x_i$ for $i = 1, \ldots, n$, and $f_{ij}(x, w)$ is the joint pdf of locations $x_i$ and $x_j$ for $i, j = 1, \ldots, n; i \neq j$. The inclusion pdf is then defined

by

$$\pi(x) = \sum_{i=1}^{n} f_i(x).$$

The joint inclusion density for $x, w \in D$ is given by

$$\pi(x, w) = \sum_{i=1}^{n} \sum_{j \neq i}^{n} f_{ij}(x, w).$$

Stevens [84] describes how the Horvitz-Thompson estimator can be used to provide estimates of the total $y_T$ over $D$ and its associated variance in terms of $y(x), \pi(x)$ and $\pi(x, w)$. An unbiased estimate of the total $y_T$ is given by

$$\widehat{y}_T = \sum_{i=1}^{n} \frac{I_D(x_i) y(x_i)}{\pi(x_i)}$$

where $I_D(x_i)$ is an indicator function that is 1 if $x_i$ is in domain $D$ and 0 otherwise. Several popular variance estimates are available, e.g. Horvitz-Thompson or the Yates-Grundy. See Stevens [84] for the details. It follows that the inclusion and joint inclusion functions determine $\widehat{y}_T$ for a probability-based design. In simple designs such as simple random sampling these inclusion probabilities are constant. In more complicated designs they can be allowed to vary according to different strata or auxiliary variables.

Similar estimates for quantities such as the mean follow naturally. For example, $\mu_t = \widehat{y}_T/|D|$, where $|D|$ represents the length of domain $D$ (i.e. the length of the river network).

### 2.4.7. Other sampling designs

Double (or two-phase) sampling offers a useful way to combine a fairly quick and cheap measurement on the quantity of interest with a more accurate but more expensive measurement technique. When the two methods are strongly (linearly) related both sources of information may be reliably brought together to provide an improved population estimate. As an example, there might be a way to measure an environmental feature rapidly and fairly exhaustively using remote sensing. When this is used in tandem with more accurate field sampling at specific sites we might be able to borrow strength from both measurements and provide improved spatial predictions of that environmental feature.

Another application of two-phase sampling is for stratification or frame development. Two-phase sampling can be very cost-effective when good frame information is difficult to develop, or stratification information is not available. Instead of needing detailed frame information for the entire population, we need it only on the first phase sample units. For example, frame information might come from high-resolution aerial photography, or even from reconnaissance visits. Such effort could be prohibitively expensive for the entire population, but feasible for the first-phase sample. Another example would be to take a first-phase sample of catchments (or watersheds), determine stream-order only for the selected catchments, and then use the stream-order to stratify. See Stevens & Jensen [91] for an application of this sampling approach.

Some environmental populations have spatial structure that makes them difficult to sample efficiently, even when employing stratagems for spatial balance. For example, natural populations frequently exhibit clustering as individuals of the same type or species tend to group together. One potential technique for improving sample efficiency for clustered populations is adaptive sampling ([92]; [93]).

Adaptive sampling allows one to modify the sample based on information as it is collected. The basic idea is best illustrated with an example. Suppose that a regular square grid has been placed over the domain of some clustered population. Further suppose that the clusters tend to be of a size that covers several grid cells i.e. the grid cell area is substantially smaller than the average cluster size. An initial sample of grid cells is selected, either by SRS or some other probability method. The cells in the initial sample are visited, and the response (e.g. the number of individual members of the target species in the grid cell) is recorded for each cell. If the response meets some pre-specified criteria (e.g. number of observed individuals is positive, or greater than some number), then adjacent cells are added to the sample. This sequence of observation/augmentation is continued until no newly observed cell meets the criteria of triggering augmentation.

The resulting sample presents some analysis difficulties, because the inclusion probability of a cell is impossible to calculate without complete knowledge of the population structure, which is not available. Thompson [91] showed how to obtain some modified weights that permitted unbiased estimates of the total using an estimator similar to the Horvitz-Thompson estimator.

There are also some difficulties in applying adaptive sampling in the field. The rule for adding to the sample must be formulated prior to beginning sampling, and must be followed in the field. In particular, new neighboring sites must be added only if the site just observed meets the criteria. Some investigators have reported that this has consequently lead to unmanageable sample sizes. Thompson [94] has recently extended the allowable stopping criteria to permit more control over the evolution of the sample. In particular, the new methodology allows the investigator to ensure a fixed sample size. However, the procedure can be computationally intensive.

Chao & Thompson [13] favour an optimal adaptive selection of sampling sites instead of an optimal conventional strategy, and demonstrate the comparison with an application to data from a study of geothermal with $CO_2$ emissions.

Model-assisted sampling is concerned with making design-based inference but using model-based estimators to improve precision ([100]). This means that sample sites are selected via a probability-based design, which provides the necessary statistical rigour for making inferences such as randomness and replication, but that auxilliary information not available through the design can be used via a model to improve precision of estimators. A comprehensive discussion of model-assisted sampling is given in Sarndal et al. [73], whilst Cressie et al. [17] discuss this approach in the broader context of accounting for uncertainty in ecological analysis.

### 2.4.8. Spatially-balanced sampling

Spatially-balanced sampling combines elements of simple random and systematic sampling. Locations are randomly selected but are guaranteed to be distributed across space in an attempt to maximise the spatial independence among sample locations. Spatially-balanced sampling has merits for sampling stream networks so we define the concept mathematically and describe a few specific spatially-balanced sampling approaches.

Yates [112] said that a sample of a response $Y$ was balanced over an auxilliary variable $Z$ correlated with $Y$ if the $z$-values (which are known beforehand) are chosen so that the sample mean of the $z$-values is exactly equal to the true population mean of $Z$. Royall & Herson [69] required the stricter condition that the first several sample moments of $z$ exactly match the corresponding population moments. The intuition behind balancing is that by forcing the $z$ sample moments to match population moments, we should get approximate balance over $Y$, and hence a more precise sample. Royall & Herson [69] show that a balanced sample is optimal in some cases.

Kott [34] noted that an option between strict balancing and SRS was to partition the range of $Z$ into quantiles, pick one point in each quantile, and then observe the corresponding $y$. While such a sample won't be strictly balanced, it does guarantee a good estimate of the distribution function of $Z$ for every sample draw. Because of the correlation between $Y$ and $Z$, one should also get a good estimate for $Y$.

If the ancillary variable is location, then we define a sample to be spatially-balanced if the spatial moments of the sample locations match the spatial moments of the population. The first two spatial moments are the center of gravity and the inertia. The center of gravity for a region $R$ is given by the ordered pair $(\mu_x, \mu_y)$, where $\mu_x$, the moment about the y-axis, is given by

$$\mu_x = \int_{-\infty}^{\infty} x v_y(x) dx.$$

The function $v_y(x)$ is the extent of the cross section of $R$ at the point $x$ and is given by

$$v_y(x) = \int_{-\infty}^{\infty} I_{\{w|(x,w)\in R\}}(y) dy,$$

where $I(\cdot)$ is the indicator function. Similar definitions hold for $\mu_y$ and $v_x$. The second spatial moment is analogous to the covariance matrix, and measures the regularity of the shape of $R$, or of the point pattern formed by the sample points.

Designs with some degree of spatial regularity or balance tend to be more efficient (i.e. yield responses that are less variable) for sampling natural resources than designs with no spatial structure. Spatial balance also ensures that there is minimal effect of spatial correlation on parameter estimates.

Some common probability-based sampling methods applied to a finite population were described and discussed in Section 2.4. Some of these designs (e.g. systematic and RTS designs) explicitly account for spatial variation and result in a spatially well-balanced sample. However all have significant limitations for large-scale environmental monitoring programs, especially for monitoring large-scale stream networks. Limitations of specific designs are mentioned in Section 2.4, but in summary, the two features of an appropriate sampling approach for monitoring stream networks which need to be accommodated are variable site inclusion probability and dynamic adjustment of the sample.

By extending the principle of the RTS design (see Section 2.4.2), Stevens [84] developed the multiple-density, nested, random-tessellation stratified (MD-NRTS) design to accommodate variable spatial sampling intensity. This design retains the spatial balance of an RTS design. Variable site inclusion probability is achieved by using a series of nested grids. The requirement for nested grids limits the choices of sampling intensity.

By generalising this process to create a potentially infinite series of nested, coherent grids, the generalised random-tessellation stratified (GRTS) design ([88]) was born. The development of this design arose through challenges in designing components of the United States' Environmental Monitoring and Assessment Program (US EMAP); this program is discussed further in Section 8.2. GRTS relies on some function that converts the population from 2-D space into 1-D space, whilst retaining proximity relationships between points in the domain. This means that an $(x, y)$ spatial address may be represented as a sequentially ordered list. Hierarchical randomisation ([87]) is used to randomly order the address and then a transformation is applied to induce an equiprobable linear structure. A sample is chosen from this randomly ordered linear list through systematic sampling. This process for sample selection is analogous to sampling a random tessellation of the 2-D space.

GRTS is an attractive design approach for sampling natural resources, including stream networks, because it

- Ensures spatial balance of sample sites prevails;
- Is flexible enough to enable dynamic adjustment of the sample size (particularly useful feature for dealing with non-response which can be substantial in stream network contexts as a result of high occurrence of dry streams, inaccessibility, lack of permission to visit a site, etc; imperfect formation of the sampling frame; and sub-populations of interest that change over time);
- Accommodates variable site inclusion probability (for dealing with potential legacy sites or for political, social or scientific reasons, sites need to be selected);
- Can be applied to monitoring of 0-, 1-, and 2-dimensional natural resources; and
- Has been successfully applied to a major natural resources (including

stream network) monitoring program in the United States for a number of years.

Design-based analysis is possible from these particular spatially-balanced designs (although model-based analyses are also an option) with further details of GRTS analytical methods discussed in detail in Diaz-Ramos et al. [20]. S-Plus and R software for both the design and analysis of GRTS samples are freely available from the US EMAP website[1].

Missing data is never handled in a pure design-based context without a model of some sort. One of the most common assumptions is that the data are Missing Completely At Random (MCAR). Under MCAR, the achieved sample is treated as if it were a SRS from the intended sample. Inclusion probabilities are adjusted by the ratio of the achieved sample size to the intended sample size. MCAR is rarely justified, but other procedures require ancillary data plus model construction. See Little [48] for a thorough background. For an environmental context, see Lesser [44]. Munoz-Hernandez et al. [60] have used a spatial model to account for missing data, whereas Stevens & Jensen [86] used post-stratification to adjust for non-response.

### *2.5. Model-based design*

Statistical models may be used to describe an underlying environmental process. Indeed, they can be used to derive an optimal spatial design through determining the number, dimensions and spatial arrangement of the sites that optimises, in some sense, the information content of the observations ([101]). These models pose important design challenges that typically manifest themselves into one of two questions: (i) how to choose an optimal design for prediction at an unknown location, and (ii) how to choose an optimal design for estimation of covariance (or variogram) parameters. These questions are however intimately related because optimal spatial prediction (kriging) relies on the spatial covariance function. If the spatial covariance function is known, the optimal spatial design is some sort of regular grid, which is preferred to a randomised sampling regime (such as simple random sampling or stratified random sampling; see Section 2.4). McBratney et al. [53] provide evidence that a triangular grid is best for minimising the average or maximum prediction variance. Cressie et al. [18] in considering spatial prediction of acid deposition, describe spatial sampling plans and optimal designs for selecting monitoring sites when there is a strong emphasis on prediction.

In order to give an appreciation for model-based approaches, and following a similar approach to Cressie [16], we consider a spatial random field given by

$$Y(x) = \mu(x) + S(x), \quad x \in D \tag{1}$$

for location $x$ and domain $D$. $\mu(x)$ is the mean process and may be a constant or described by some spatial trend, e.g. $\mu(x) = \beta' \alpha(x)$, where $\alpha(x)$ is a vector of

---

known functions of location $x$, e.g. polynomials, and $\beta$ is a vector of coefficients. $S(x)$ is a spatial random effect such that

$$E[S(x)] = 0$$

and

$$Var[S(x), S(x^{'})] = \sigma^2 \rho(\parallel x - x^{'} \parallel, \theta)$$

for some isotropic correlation function $\rho$ with covariance parameters $\theta$. In the prediction problem at this stage we assume that these parameters are known. This spatial dependence is often represented in terms of the variogram $2\gamma(\cdot)$ or semi-variogram $\gamma(\cdot)$. For a stationary process, the semi-variogram can be defined as

$$\gamma(\parallel x - x^{'} \parallel) = \sigma^2(1 - \rho(\parallel x - x^{'} \parallel, \theta)).$$

The optimal spatial prediction for a new location $x_0$ is given by the kriging or best linear unbiased prediction (BLUP) estimator, namely

$$\widehat{Y}(x_0) = \sum_i \lambda_i Y(x_i)$$

where $\lambda$ are the kriging weights and are described in any textbook on geostatistics such as Cressie [16]. The kriging variance associated with location $x_0$ is given by

$$\sigma^2(x_0) = 2 \sum_{i=1}^{n} \lambda_i \gamma(x_i - x_0) - \sum_{i=1}^{n} \sum_{j=1}^{n} \lambda_i \lambda_j \gamma(x_i - x_j)$$

for semi-variogram $\gamma$. Note that this kriging variance depends on the number of points, the position of these points and the variogram/covariance function. There is no dependence on the actual values at those points.

In all but the simplest cases it is not feasible to choose the optimal position of the design points without applying some restriction to narrow the choice. The first of these is to discretise domain $D$ so there are a finite number of possible locations. This is however typically not enough. If $N$ is the total number of potential locations and $n$ is the number that we want to select, ${}^N C_n$ may still be prohibitively large for even a modest choice of $N$ and $n$.

In an iterative design we look to choose a new location $x_0$ given current design points $x_1, \ldots, x_n$. A common criterion for doing this is to minimize the average (or maximum) prediction or kriging variance. This means we choose $x_0$ to minimise

$$V(x_0) = \int_D \sigma^2(x, x_0; x_1, \ldots, x_n, \theta) \, dx.$$

This integral is usually impossible to integrate directly. Instead the domain $D$ is described by a discrete set of new or potential locations and

$$V(x_0) \approx \sum_x \sigma^2(x, x_0; x_1, \ldots, x_n, \theta)$$

is used as the criterion to minimise. This can be extended by introducing a weight function $w(x)$ to influence where the new sample locations are placed, i.e.

$$V(x_0) = \int_D \sigma^2(x, x_0; x_1, \ldots, x_n, \theta) w(x) \ dx.$$

If $w(x) = 1$, no restrictions are placed. Alternatives may force the observation to occur in a region $R$ by using $w(x) = I(x \in R)$ or be placed at hotspots in terms of the mean, such as $w(x) = I(\mu(x) > K)$ or variance, such as $w(x) = I(\sigma^2(x) > L)$. Note that $I(u)$ is an indicator function with value 1 if $u$ is true and 0 otherwise. Combinations of indicator functions are also possible.

This iterative design, where points are added sequentially so as to optimise a criterion like the average prediction variance, will not (as a default) yield the same set of design points as something that optimises over all design points simultaneously. It should however be viewed as delivering approximate optimality in a manner that will be computationally more feasible. Extensions are possible where $x_0$ is chosen to be optimal for block prediction variance. Other variants of kriging might also be used.

Regular grids are not optimal if the goal is estimation of the covariance function because it is important to have locations that are closer together to allow the estimation of the short range spatial variation with any reliability. Stein [82] shows that the kriging estimates are most sensitive to error in the variogram at short ranges. Warwick & Myers [104] consider the design problem for variogram estimation and focus on the distribution of lags (and directions/angles). Bogaert & Russo [5] and Muller & Zimmerman [59] consider a generalised least squares fit to the empirical variogram and use features of the covariance matrix of the variogram parameters as design criteria. Lark [38] and Zhu & Stein [113] use likelihood methods to select the optimal design for parameter estimation.

For a spatial process of the form (1) with $\mu(\beta, x) = \beta' \alpha(x)$, the log-likelihood of $Y$ is given by

$$l(\beta, \phi, x_1, \ldots, x_n) \propto -\frac{1}{2} \log \det \Sigma(\phi) - \frac{1}{2}(Y - \mu(\beta, x))' \Sigma(\phi)^{-1}(Y - \mu(\beta, x))$$

where $\phi = [\sigma^2, \theta]$. Since the covariance matrix of $\phi$ quantifies our uncertainty about $\phi$ given the data, the design criteria might focus on the asymptotic covariance matrix as given by the inverse of the Fisher information matrix. Zhu & Stein [113] consider the following criterion

$$V(\phi, x_1, \ldots, x_n) = -\log \det I(\phi, x_1, \ldots, x_n)$$

where $I(\phi, x_1, \ldots, x_n)$ is the Fisher information matrix, and is defined as

$$I(\phi, x_1, \ldots, x_n) = E_\phi \left[ \frac{\partial}{\partial \phi} l(\beta, \phi) \left\{ \frac{\partial}{\partial \phi} l(\beta, \phi) \right\}' \right].$$

This formulation uses the determinant as a generalised measure of uncertainty in $\phi$.

If the objective is to satisfy both prediction and the estimation of the covariance function, the spatial design is typically a combination of some sort of regular grid with additional points at shorter distances. For instance, Xia et al. [111] use likelihood-based methods and focus on design criteria based on the Fisher information matrix that consider both the mean and covariance structure in an application to understand the distribution of Toxic Release Inventory (TRI) chemicals released into the environment. Zhu & Zhang [114] look at optimal spatial designs for prediction with estimated parameters using infill asymptotics to reparameterise the covariance functions and simplify the design criteria. They note that the choice of design criteria is a non-trivial task. The choice they settle on amounts to a linear combination of criteria that are reasonable for each objective. Marchant & Lark [51] is a recent paper which discusses how to optimize sampling schemes for geostatistical surveys when both variogram estimation and kriging (prediction) are of interest.

Zimmerman [116] notes that the design objectives for efficient prediction assuming known dependence and efficient estimation of spatial dependence parameters are largely antithetical and often lead to very different optimal designs. Zimmerman introduces a hybrid design that emphasises prediction but accounts for the uncertainty in the covariance parameters. His approach is to choose the design to minimise an approximation to the variance of the empirical kriging (empirical-BLUP) prediction error. Note that the empirical kriging/BLUP predictor involves evaluating the covariance matrix at the estimated $\hat{\theta}$ rather than the assumed known $\theta$. He applies his findings to an acid deposition data set from the eastern USA.

Diggle & Lophaven [21] describe a Bayesian approach to spatial design that balances the design for parameter estimation with spatial prediction. The designs are efficient for spatial prediction but make an appropriate allowance for parameter uncertainty. They also compare the efficiency of designs based on a regular grid plus extra close pairs to a regular grid with in-filling. An application to reducing sampling intensity for measuring surface salinity at monitoring stations in the Kattegat Basin, Scandinavia, is described.

Diggle & Lophaven [21] note that under a Bayesian paradigm, the predictive distribution at location $x$ is given by

$$p(y(x)|y) = \int p(y(x)|y; \phi) \ p(\phi|y) \ d\phi \qquad (2)$$

where $y = y(x_1), \ldots, y(x_n)$. This is a weighted average of the classical predictive distribution $p(y(x)|y; \phi)$ where the weights are assigned according to the prior distribution for $\phi$. Diggle & Lophaven [21] use a design criterion that is based on the average prediction variance over the region, that is

$$V(x_1, \ldots, x_n) = \int_D var(y(x)|y) dx. \qquad (3)$$

Various candidate designs, as defined by their spatial locations $x_1, \ldots, x_n$ are considered. For each design the average prediction variance is calculated from the predictive distribution $p(y(x)|y)$. This is typically calculated using Monte Carlo methods. A sample value $\phi$ is generated from $p(\phi|y)$. This sample value is then attached to the $p(y(x)|y; \phi)$, and sampled from this to obtain realisations of the predictive distribution. These steps are repeated many times to generate a sample from the predictive distribution in (2). The spatial integral in the average prediction variance criterion in (3) is calculated by replacing the integration over continuous domain $D$ by a summation over discretised domain points $x$. Diggle & Lophaven [21] find that a wide range of distances should be included in the spatial design, and that the popular regular design is often not the best choice.

There are significant challenges in optimising the location of monitoring sites for a spatial design. It is invariably necessary to discretise potential monitoring locations and choose amongst that set. If the possible sites $N$ and the number of sites to choose $n$ are small it may be feasible to enumerate all possibilities. For even a modest number of sampling locations there is a prohibitively large space over which to optimally choose all locations simultaneously. No algorithm is currently available to find an exact solution in reasonable time ([113]). Simulated annealing has been a popular combinatorial optimisation algorithm for obtaining a near optimal solution ([23]; [99]; [113]).

It is often possible to achieve approximate optimality, by adding sites sequentially, with the site that reduces the average prediction variance (kriging variance) by the most being selected at that time. Computational feasibility may also be achieved by considering different families of designs and identifying the optimal spatial design amongst all members of the family rather than satisfy the unconstrained problem.

Model-based design has also featured strongly in entropy-based design. This typically involves choosing sites to somehow maximize information content. Sites are added or subtracted to minimize entropy, where low entropy corresponds to high information content. There are many papers which consider optimum spatial sampling design based on entropy, including Angulo et al. [1], Bueso et al. [8], Casselton & Zidek [11], Casselton et al. [12], Fuentes et al. [26], Le & Zidek [42], Le et al. [43], and Zidek et al. [115]. Under the assumption of Gaussian random fields, the optimum choice of an additional site turns out to be the site with the largest conditional variance (i.e. variance conditional on the sites already selected in the design). The majority of these authors demonstrated their methodological developments on applications concerning atmospheric variates such as $PM_{10}$. The entropy-based approach can be motivated using Bayesian arguments.

The theory of optimal experimental design has also had an important impact on model-based design. This often comes under the collective heading of *optimal design theory* and relates to choosing the design points according to their ensuing efficiency for estimating regression parameters. This appeals to a classical theory of optimal design and seeks to choose the design matrix over all design matrices to optimise this criterion. Design concepts such as D-optimality and A-optimality are important as the criterion is usually based on the information matrix. Fedorov [25] contributes an important review paper in this field. Muller [57] provides a comprehensive account.

Optimal experimental design theory is not that commonly applied to correlated spatial data because it is not that straightforward when the assumption of independent errors is violated. Müller [58] gives an account of several possibilities for dealing with this issue, and demonstrates this comparison through application to the design of a water-quality monitoring network. Smith [79] provides an excellent summary of entropy and optimal design theory approaches in his online notes.

Wiens [107] discusses robust designs for spatial processes in the face of uncertainty about measurement errors and the spatial covariance. A minimax approach is adopted. This uses a loss function based on the mean-squared prediction error that is maximised over departures from the fitted regression response. Sites are then selected to minimise the maximum loss according to a simulated annealing algorithm. A coal-ash data set are used to illustrate this approach.

## 3. Informing the design using expert elicitation

The spatial design for natural resource monitoring applications, particularly large-scale monitoring programs, is often limited by the availability of relevant historical data. Consequently, input from "experts" is usually required to identify suitable sites for inclusion into the sampling regime that satisfy logistical constraints. This is often referred to as soft data, and is particularly valuable when relevant "hard" data (i.e. actual observations) and knowledge of the domain are limited. The method of extracting relevant and important information from experts (either a single expert or multiple experts) is referred to as expert elicitation. Eliciting prior information and its use is a well-researched area in psychology however it is a relatively new research area in the environmental sciences and statistics. Lele & Allen [42], Lele & Das [43], Martin et al. [54] and Kynn [36] discuss recent research on expert elicitation for ecological applications, whilst Baddeley et al. [2] consider incorporating prior information in geoscience contexts.

In a design context, expert elicitation has been undertaken in a fairly informal and ad hoc manner to date and there is a need to formally recognise the importance of this step in the overall monitoring framework. There are two ways of using expert elicitation to inform the spatial design:

1. use it to derive an inclusion probability for a site from the target population being selected for the required sample, based on a combination of logistical constraints, historical information and knowledge; or
2. use it to derive inclusion probabilities for a site from an inflated sample being selected for the required sample, based on a combination of logistical constraints, historical information and knowledge.

In either case, the inclusion probabilities will be less subjective and more informative when selecting sample sites from a population than those derived simply by adhering to the specific adopted design.

Bayesian methods are more commonly used to incorporate the elicited information into a model framework through probability statements (for example see [36] or [52]), however frequentist approaches have also been considered for incorporating this information; see Lele & Allen [42]. Differences in the approaches come down to whether the elicited information is in the form of a prior distribution (or parameters of that distribution) or in the form of data.

Common methods for elicitation include questionnaires, a simultaneous meeting of experts (expert panel), making use of a relevant geographical information system (GIS), or using a combination of these (i.e. adaptive approach). An example of using a combination of methods is extracting a prior or data from an expert (either via a questionnaire or expert panel) and then using a GIS-based model to derive more information about conditions to refine the prior/data. Methods may be direct (elicitation of the response of interest) or indirect (elicitation of an explanatory variable). To ensure expert responses are valid and consistent, it may be beneficial to give them a discrete choice of answers to questions e.g. "does turbidity increase, decrease or stay the same, under particular conditions?" It may also be of benefit to use professional elicitors/mediators to extract information from the experts to minimise bias of that information.

## 4. Spatio-temporal designs

Environmental monitoring often aims to provide information that allows us to determine if there has been any change over time in important environmental variables (e.g. % of stream with low dissolved oxygen). This may be particularly relevant following a significant event (e.g. a major storm) or some management intervention (e.g. the reduction in nutrient output from a waste water treatment plant following the adoption of improved practices). Spatial designs often need to be repeated over time. It is important that the spatial and temporal components be considered simultaneously in order to ensure effectiveness.

The temporal component of the monitoring design is clearly fundamental if we are to detect any temporal changes or trends[2]. This often begins with the need for a strong understanding of the current and baseline condition if we are to detect any future change[3]. For this reason it is often necessary to invest greater monitoring resources at the beginning of a monitoring program to ensure these baseline conditions are adequately measured and understood and the spatial and temporal variability in the environmental quantities of interest are quantified at all relevant scales.

Environmental monitoring designs over time are more complicated than single time surveys as there is a need to consider both spatial and temporal designs simultaneously. Lettenmaier [45] describes temporal design considerations for ambient stream quality monitoring whilst Overton & Stehman [74] describe some of the desirable design characteristics for long-term monitoring of ecological variables more broadly. If relevant historical data can be used to quantify temporal correlations (e.g. an AR(1) process may adequately describe the correlation between monthly flow inputs to a stream network) then this would be of great benefit to informing the specificity of the temporal design.

---

[2]We define temporal trend as the change in a series of data over a period of years that remains after the data have been adjusted to remove known effects such as season, long-term cycles and other relevant covariates.

[3]Change can be absolute (or gross), that is, the comparison of a current measurement to a prior measurement, or it can be relative (or net), that is, the comparison of a current measurement to a prior measurement, after taking into account the comparative shift in the baseline measurements.

TABLE 2
*A pure panel design.*

| Panel or site set | Time | | | | |
|---|---|---|---|---|---|
| | $y_1$ | $y_2$ | $y_3$ | $y_4$ | $y_5$ |
| 1 | X | X | X | X | X |

TABLE 3
*An independent samples design.*

| Panel or site set | Time | | | | |
|---|---|---|---|---|---|
| | $y_1$ | $y_2$ | $y_3$ | $y_4$ | $y_5$ |
| 1 | X | | | | |
| 2 | | X | | | |
| 3 | | | X | | |
| 4 | | | | X | |
| 5 | | | | | X |

There are a variety of design possibilities available; Fuller [27] and McDonald [54] provide indepth accounts of the options, whereas Urquhart et al [97] present a comparison of two classes of sampling designs for long-term environmental monitoring.

Panel designs have been common place in the survey sampling literature for a long time ([77]) and now feature prominently in environmental monitoring design ([27]; [54]). 'Panel' in the environmental monitoring context refers to a collection of monitoring sites.

In the context of drawing 'panels' of spatially-balanced probability samples, we note that any set of consecutively-numbered sample sites in a GRTS sample is in itself a spatially well-balanced sample ([88]). Consequently, an inflated GRTS sample, also known as an oversample, could be drawn from a population and consecutive sets of sample sites could be used as panels in long-term monitoring program. An example for such a use is in drawing panels of interpenetrating sites that may represent consecutive years in a monitoring program for assessing change in condition of a region.

Some of the basic panel designs include:

- *Pure panels* where a single sample of sites is chosen and repeated each year, as displayed in Table 2.

- *Independent samples each year* where a different set of sites are sampled each sampling run, as illustrated in Table 3.

- *Rotating panel design* where a set of samples are not all visited each year.

TABLE 4
*A rotating panel design*

| Panel or site set | Time | | | | | |
|---|---|---|---|---|---|---|
| | $y_1$ | $y_2$ | $y_3$ | $y_4$ | $y_5$ | $y_6$ |
| 1 | X | | | X | | |
| 2 | | X | | | X | |
| 3 | | | X | | | X |

TABLE 5
*An alternative rotating panel design*

| Panel or site set | Time | | | | |
|---|---|---|---|---|---|
| | $y_1$ | $y_2$ | $y_3$ | $y_4$ | $y_5$ |
| 1 | X | | | | |
| 2 | X | X | | | |
| 3 | | X | X | | |
| 4 | | | X | X | |
| 5 | | | | X | X |

In addition, different panels may have different sampling schedules. Tables 4 and 5 display two different rotating panel designs. The former is called a *serial alternating design* by [97].

- *Split (also supplemental or augmented) panel designs* which have a set of pure panel samples and a set of random independent samples, as displayed in Table 6.

The optimal panel design depends on the balance between needing to detect trend and report on status/condition. Designs that revisit sites are more efficient for estimation of trend, while designs with independent sites are generally more efficient for estimation of status. Split panels are generally thought to provide the best chance of satisfying competing objectives of a monitoring program ([54]). Figure 2 gives a representation of the relationship between the monitoring objectives and the ideal panel method.

There has been some work on the power of various designs to detect change; see Urquhart & Kincaid [96] and Urquhart et al. [97, 98].

TABLE 6
*A split panel design*

| Panel or site set | Time | | | | |
|---|---|---|---|---|---|
| | $y_1$ | $y_2$ | $y_3$ | $y_4$ | $y_5$ |
| 1 | X | X | X | X | X |
| 2 | X | | | | |
| 3 | | X | | | |
| 4 | | | X | | |
| 5 | | | | X | |
| 6 | | | | | X |

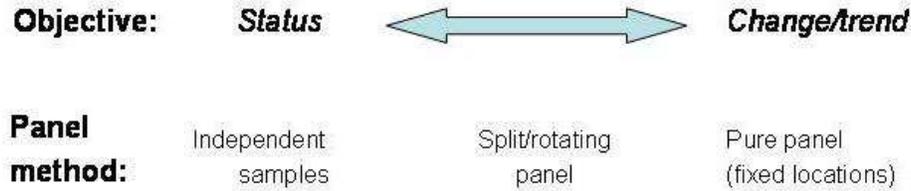

Fɪɢ 2. *Panel designs and the emphasis on status vs trend detection.*

Some environmental or ecological processes will change spatially over time. For instance, some wildlife species may move according to available food sources. Typically processes in streams are less dynamic than air but more dynamic than soil. If static spatial designs are used, the same spatial locations are visited over time and there may be some inefficiency in the use of monitoring resources because they may not always be placed in the best place to detect change or report on status. In addition, there is serious potential to impact the site through repeated visits to particular sites, in which case the environment may alter and inferences may not be comparable to those made from initial visits to those sites.

Wikle & Royle [108, 109] describe spatially dynamic designs for ecological and environmental monitoring contexts. These designs recognise that ecological processes change over time in a coherent, dynamic fashion and that by considering the joint spatio-temporal dependence the designs will choose locations to provide the most information. Spatially dynamic designs are more complicated than static spatial designs, and are more important when there is stronger temporal correlation as that gives us the best ability to predict the ecological process in the future from current conditions and place monitoring resources appropriately. Wikle & Royle [108] describe spatially dynamic designs for Gaussian processes, while Wikle & Royle [109] extend these designs to the more complicated case of non-Gaussian data.

## 5. Other important design considerations

When it comes down to it, money is ultimately the biggest factor that drives the design. Costs related to sampling a suite of indicators, including materials, equipment, transport, laboratory analysis, and the time of trained staff, often dictate the number of sites that can be sampled and the information that may be collected from each site. With the advent of sensor networks (discussed in Section 6), we acknowledge that some of these costs may be less of an issue in the future.

Other important practical ways of maximising the information available for a fixed cost include considering how to:

- minimise non-response, which, for example, may occur if a site happens to be located in a 'dry' stream thus making a water quality assessment impossible to undertake;

- ensure accessibility of sites (e.g. inaccessibility may occur due to lack of permission to access a property or stream reach or due to physical impediments);
- ensure quality of data processing (i.e. from recording of the data at the site, perhaps by electronic means, to data checking and cleaning, through to statistical analysis and reporting).

The design will meet the pre-defined (usually multiple) objectives of the study. The sample design, being just one component of a monitoring program framework, is directly linked to and driven by those monitoring objectives. Different objectives will necessarily lead to different spatial and temporal designs. For example, a design to assess ambient water quality condition will be quite different to a design which tries to meet objectives concerning impacts of a particular event on the water quality.

If assessing condition is of interest then relevant *reference* sites (typically sites that have had no or little impact) need to be selected as part of the sample too. Random selection of such sites may not be possible because it is difficult to identify a reliable target population of reference sites. It is also possible that reference sites in a system may not exist, either due to the extent of human activities, or prevailing climate conditions (e.g drought),

The choice of appropriate indicators to sample can be non-trival for monitoring programs concerning assessment of stream condition. The number and specific nature of these indicators (such as their components of variance, and their scale of application) should be taken into account when developing an optimal spatial design.

An important but often overlooked concept associated with design is statistical power. Power is an outcome of a statistical test of the resulting data; for environmental contexts it is often defined as the probability of correctly concluding that there was an impact (or difference) in a study. Whilst power is subjective and needs to be controlled by the researcher, there is a trade-off between monitoring costs, precision of estimates and the power to correctly identify when no impact has occurred. It is related to the effect size (i.e. a meaningful ecological difference), the sample size of the design and the significance level for statistical tests, in that the greater they are, the greater it is, for any pairwise comparison holding the other two design parameters constant. It is recommended that the power of a test be as large as possible to minimise risk of a Type II error[4] occurring. For further discussion of statistical power in an environmental context, see [24].

When designing a monitoring program, it is essential to consider the data that will result from implementing the design and how these data will be analysed to address the objectives and communicate outcomes of the program. For example, we may want to choose sites which are correlated to borrow strength in forming predictions i.e. choosing a certain proportion of sites to inform short-range and long-range variation. Such selection of correlated or dependent sites contradicts the classical design principle of selecting (spatially-)'independent' samples. The main benefit in choosing spatially-dependent sites is to attempt to quantify and explain the sources of spatial variation for a particular response, namely its spatial covariance, which will enable more confident predictions of that response at unmonitored sites in the region.

---

[4]The probability of concluding that there was no impact when really one did exist. It is also known as a false negative error.

A distinction should be made between a sample design and a response (or operational) design. The former is used to select the sample sites (the *where* and *when*) and the latter to prescribe how responses are collected (the *how*, *what* and *by whom/what*), although the two are intrinsically related. In the context of monitoring the condition of streams, a response design will help quantify measurement error and validate protocols written to guide field workers who undertake the physical sampling and data collection and input. In particular, it should help reduce possible bias that may otherwise arise e.g. due to purposive (also known as convenience or haphazard) ambient monitoring.

Exclusion zones are areas in the target sample frame in which sample locations are given zero probability of inclusion in the resulting sample. These are often present to reduce the search area and make better use of sampling resources. It does however mean that no inference can be extended to these zones. This can be problematic because

- knowledge / expectations about these zones is not always correct which leads to incorrect exclusions;
- surrogate data used to generate these exclusion zones are subject to uncertainty; and
- the variability of natural processes often means there will be change in the process over time.

Theobald et al. [90] argue that it is better to allocate a modest amount of resources to low priority areas than to exclude them. For natural resource applications, exclusion zones may arise due to areas of definite "no go" (e.g. indigenous areas of cultural importance) or due to recommended agency monitoring protocols such as that sampling needs to be carried out within a prespecified distance from the nearest road.

## 6. Computational advances in monitoring and design

In the majority of cases, (large-scale) monitoring programs are these days efficiently designed using appropriate software, statistical or otherwise. Most statistical packages have design components and capabilities, especially for some of the standard probability-based designs. And as Theobald et al. [90] point out, most standard Geographic Information Systems (GIS) provide some tools to construct spatial survey designs. However when it comes to non-standard or complex spatial designs, customised programs or scripts will be required. They advocate and discuss in detail several advantages of generating a spatially-balanced design within a GIS framework, namely

1. A GIS is typically used to construct the target population from which to draw a sample;
2. A GIS enables visualisation of a sample design in relation to other demographic or geographic features in the target region such as a road network (which may be important for finetuning the sample and response design); and
3. A broader base of potential users will be reached.

In this age of "Everything, Everywhere" ([9]), sensor networks are fast becoming of increasing importance and practicality in relation to monitoring design and adaptation of design. They will offer advances in real-time (i.e. on-line) monitoring and are providing new statistical research directions with respect to design (sampling and response) and analysis for large-scale programs. If sensor networks do form part of a design, it is highly recommended to conduct a pilot study on a small subset of the intended nodes in order to iron out potential issues arising from their use in field that are not predictable on paper. For instance, there may be unique challenges posed by the nodes only being able to sense for limited durations due to power constraints that will affect the spatial configuration. It may also be necessary to consider how to handle the challenge of the large amount of data generated by sensor networks in the design phase.

The increasing availability of high performance computing capability and capacity is promising for aquatic monitoring design as these facilities offer ability to deal with ensuing big data sets. Indeed many spatial analysts already use such facilities to analyse their data in order to make use of modern computational methods such as Markov Chain Monte Carlo (MCMC) and for dealing with large matrix calculations or to perform intensive optimisation tasks like simulated annealing.

Some of the big computational challenges that lie ahead concern implementation of a design which may require marriage of different technologies for different scales e.g. a GIS used to generate a digitised target population, statistical software is subsequently used to generate a (inflated) set of sample sites, followed by expert elicitation which is used to subset those sample sites where there actually is water to sample (which may take the form of a probability surface for a region or may be a presence/absence response for a particular sample site). These kinds of "meta-designs" may become more commonplace as statisticians strive to design programs that meet multiple objectives, are large-scale, are guided by multi-disciplinary teams, and are dealing with complex systems.

## 7. Applications of spatial design for monitoring stream networks

In the past 20 years, the number of publications which present innovative spatial designs for monitoring stream networks or discuss applications in the broader context of aquatic monitoring has been on the increase, largely due to the publication of research directly related to, or stemming from, the United States Environmental Monitoring and Assessment Program (US EMAP). However there have been a modest number of publications from applications of assessing condi-tion of Australian streams. We briefly consider a few of these Australian stream network applications and summarise their spatial design components. A succinct historical account of the US EMAP is subsequently presented.

Dixon et al. [23] described a method to assist managers with optimising the selection of river sampling sites, to provide a justifiable, best choice design for single objective programs or a best compromise design for hybrid monitoring programs or those with constraints. Their method combined GIS, graph theory and the simulated annealing algorithm. They illustrated this method with three case studies: a simple regulatory monitoring situation; a situation where possible sampling sites are severely restricted; and monitoring an impounding catchment with problem inflows. Each of these case studies was undertaken in South-east Queensland, Australia.

In 2001, Smith & Storey [78] published a report that summarised outcomes of development of a cost-effective, coordinated Ecosystem Health Monitoring Program (EHMP) for freshwaters in southeast Queensland. The general aim of this program is to measure and report on current and future changes in ecological health of this system under ambient conditions. As part of the design, a classification of southeast Queensland's streams was undertaken for multiple reasons including aiding site selection and allowing sites to be stratified across all the major stream types within the study area. One of the objectives of the pilot study was to evaluate the utility and variability of less-proven indicators for assessing ecological health of rivers. So a major focus of the pilot was a field evaluation of various methods across a disturbance gradient (land clearing). Recommendations on the spatial scale of the ambient EHMP were to allocate sites to each (Strahler) 3rd-order stream and to the catchments of all larger (Strahler) 2nd order streams until adequate spatial coverage had been achieved. This was on the basis that 1st, 2nd and 3rd-order streams comprise nearly 90% of stream length in southeast Queensland.

The Sustainable Rivers Audit ([80]) arose from recognising the need for a comprehensive, basin-wide monitoring program to provide vital information on the overall health status of the Australian Murray River, the Darling River and their major tributaries which together form the largest freshwater basin in Australia. The basin-wide audit considers numerous indicators of fish, macroinvertebrates and hydrology of regulated rivers at various scales with the aim of assessing river health. A statistically robust method of site selection was developed. An example of a possible site selection method - stratified random sampling - is described in detail in Appendix 2 of SRA [80]. As a result of conducting a pilot audit, several issues arose:

- There were a high number of 'dry' sites. They hope to overcome this through more rigorous selection of the targeted stream network before site selection is undertaken.
- For those catchments where nearly all streams could be classified as ephemeral, they may aim to sample during a 'wet' year.
- The random site selection method resulted in several sites being clumped

together and large areas of catchments not being sampled at all. To overcome this issue, a couple of distance-related rules were suggested.

They also advocated that once a site had been randomly selected, it should be sampled regardless of its perceived condition. Sites are seen as replicates and not as being representative of the condition of the region.

In response to concerns over acidification of surface waters, the United States Environmental Protection Agency (USEPA) initiated the National Surface Water Survey (NSWS) in 1983, providing status reports on lakes ([47]; [37]) and streams ([55]; [33]). These surveys were based on probability samples of lakes and streams in sensitive areas of the United States. Although large-scale, probability-based surveys of natural resources had been applied earlier for forests ([30]) and agriculture ([15]), the NSWS were the first nation-wide, probability-based surveys of an aquatic resource.

The success of the NSWS demonstrated the utility of probability samples in determining the status of well-defined resource populations. In 1988, the USEPA's Science Advisory Board recommended a program within EPA to monitor status and trends of the condition of ecological resources and to develop innovative methods for anticipating emerging environmental problems before they reach crisis proportions. The Environmental Monitoring and Assessment Program (EMAP) was initiated by EPA's Office of Research and Development to respond to this recommendation and generally to provide a greater capacity for assessing and monitoring the condition of the nation's ecological resources ([35]; [56]).

EMAP was initially based on the concept of a random tessellation stratified design ([62]; [66]) using a hexagonal tessellation. The tessellation was based on a global grid derived from a truncated icosahedron ([106]). The icosahedron was positioned so that the continental United States was covered by one of the hexagonal faces of the icosahedron, and then given a small random displacement. The original EMAP design is described in Overton et al. [68].

The basic RTS design of EMAP was extended to allow some variable probability designs and nested designs by Stevens [84]. These designs were applied to Lake Ontario to measure toxicity in sediment and water column ([49]), streams in the Mid-Atlantic region of the US ([31]), and to the Southern California Bight (SCB) ([3]). The SCB design used (1) unequal probability selection based on overlapping subpopulations that were of interest, and (2) nested subsampling of indicators related to increased cost to acquire some indicators.

The nested, variable-probability design used for the SCB spurred the development of an even more general design paradigm, the Generalized Random Tessellation Stratified design ([87, 88]). Stevens [85] gives an example of the GRTS design applied to monitoring Coho salmon in coastal streams of Oregon, USA. Wardrop et al. [102, 103] and Whigham et al. [105] discuss the results of a GRTS design applied to wetlands in two large basins in the USA.

The GRTS design has been applied by many state agencies in the USA. A partial list with accompanying design documentation can be found on the US EPA website[5]. In addition, GRTS has been applied to sampling rivers and streams in Portugal (Personal communication, Tony Olsen, USEPA).

## 8. Conclusions

The development of appropriate spatial designs for monitoring stream networks to meet multiple objectives is typically a non-trivial task hindered by the complexity and scale (usually landscape scale) of the system being monitored and the practical challenges faced in sampling these types of systems. This review has attempted to consider and synthesise the large body of literature on approaches to spatial design for the specific context of monitoring stream networks, that we refer to as sparse sampling. We have documented the current status of this area of research, with the aim of helping determine the most practical and optimal spatial design approach for stream network domains.

Interestingly, in most of the literature we considered (not all of which was cited here), developments in model-based design approaches were motivated by applications in two-dimensional domains such as air, estuaries and marine environments, and soil. Few were motivated by linear systems such as streams. On the other hand, applications in monitoring large-scale natural resource systems including stream networks have motivated some recent developments in probability-based design.

One of the outcomes of this review is that probability-based design approaches appear to be better suited both theoretically and practically for the sparse sampling of stream networks. Common approaches such as simple random sampling and stratified sampling have limitations when it comes to sampling large-scale stream networks, but one such promising approach is the generalised random-tessellation stratified (GRTS) sampling design ([88]). The strengths of this approach for monitoring stream networks are numerous, but include ensuring spatial representation of sites, enabling the sample size to be dynamically adjusted to help minimise non-response, and allowing sites in the population to have variable probability of inclusion in the sample.

## Acknowledgements

We thank the Associate Editor and two referees for suggestions that helped improve the clarity, structure and focus of the paper.

---

[5]http://www.epa.gov/nheerl/arm/programpages/state_strategy.htm